\documentclass[twocolumn,showpacs,superscriptaddress,amsmath,amssymb]{revtex4}
\usepackage{graphicx}
\usepackage{dcolumn}
\usepackage{bm}
\usepackage{color}
\usepackage{mathrsfs}
\usepackage{ulem}

\def\bd{
\begin{document}} \def\ed{\end{document}}
\def\bmp{\begin{minipage}} \def\emp{\end{minipage}}
\def\bcc{\begin{center}} \def\ecc{\end{center}}     \def\npg{\newpage}
\def\beq{\begin{equation}} \def\eeq{\end{equation}} \def\hph{\hphantom}
\def\be{\begin{equation}} \def\ee{\end{equation}} \def\r#1{$^{[#1]}$}
\def\n{\noindent} \def\ni{\noindent} \def\pa{\parindent}
\def\hs{\hskip} \def\vs{\vskip} \def\hf{\hfill} \def\ej{\vfill\eject}
\def\cl{\centerline} \def\ob{\obeylines}  \def\ls{\leftskip}
\def\underbar#1{$\setbox0=\hbox{#1} \dp0=1.5pt \mathsurround=0pt
   \underline{\box0}$}   \def\ub{\underbar}    \def\ul{\underline}
\def\f{\left} \def\g{\right} \def\e{{\rm e}} \def\o{\over} \def\d{{\rm d}}
\def\vf{\varphi} \def\pl{\partial} \def\cov{{\rm cov}} \def\ch{{\rm ch}}
\def\la{\langle} \def\ra{\rangle} \def\EE{e$^+$e$^-$} \def\pt{p_{\rm t}}
\def\pti{p_{{\rm t},i}} \def\vti{v_{{\rm t},i}}
\def\ptj{p_{{\rm t},j}}\def\Pt{P_{\rm t}} \def\vt{v_{\rm t}}

\def\bitz{\begin{itemize}} \def\eitz{\end{itemize}}
\def\btbl{\begin{tabular}} \def\etbl{\end{tabular}}
\def\btbb{\begin{tabbing}} \def\etbb{\end{tabbing}}
\def\beqar{\begin{eqnarray}} \def\eeqar{\end{eqnarray}}
\def\\{\hfill\break} \def\dit{\item{-}} \def\i{\item}
\def\bbb{\begin{thebibliography}{9} \itemsep=-1mm}
\def\ebb{\end{thebibliography}} \def\bb{\bibitem}
\def\bpic{\begin{picture}(260,240)} \def\epic{\end{picture}}
\def\akgt{\cl{\bf ACKNOWLEDGMENTS}}
\def\fgn{\noindent{\bf\large\bf figure captions}}
\def\m1{\langle N_p\rangle} \def\u2{\langle N_{\bar p}\rangle} \def\Nap{N_{\bar
p}}
\def\lan{\langle}
\def\ran{\rangle}
\def\p{\pi}
\def\ifmath#1{\relax\ifmmode #1\else $#1$\fi}%
\def\rc{\ifmath{{\mathrm{c}}}}
\def\cut{\ifmath{{\mathrm{cut}}}}
\def\rF{\ifmath{{\mathrm{F}}}}
\def\rK{\ifmath{{\mathrm{K}}}}
\def\rp{\ifmath{{\mathrm{p}}}}
\def\rt{\ifmath{{\mathrm{t}}}}
\def\LAB{\ifmath{{\mathrm{LAB}}}}
\def\cut{\ifmath{{\mathrm{cut}}}}
\def\beq{\begin{equation}}
\def\eeq{\end{equation}}

\newcommand{\cinst}[2]{$^{\mathrm{#1}}$~#2\par}
\newcommand{\crefi}[1]{$^{\mathrm{#1}}$}
\newcommand{\crefii}[2]{$^{\mathrm{#1,#2}}$}
\newcommand{\crefiii}[3]{$^{\mathrm{#1,#2,#3}}$}
\newcommand{\HRule}{\rule{0.5\linewidth}{0.5mm}}

\bd
\title{ Poisson baseline of net-charge fluctuations in the relativistic heavy ion collisions}

\author{Xue Pan}\email{panxue1624@163.com} 
\affiliation{School of Electronic Engineering, Chengdu Technological University, Chengdu 611730, China}
\author{Yufu Lin}
\affiliation{Key Laboratory of Quark and Lepton Physics (MOE) and Institute of Particle Physics, Central China Normal University, Wuhan 430079, China }
\author{Lizhu Chen} 
\affiliation{ School of Physics and Optoelectronic Engineering, Nanjing University of Information Science and Technology, Nanjing 210044, China}
\author{Mingmei Xu} 
\affiliation{Key Laboratory of Quark and Lepton Physics (MOE) and Institute of Particle Physics, Central China Normal University, Wuhan 430079, China  }
\author{Yuanfang Wu} 
\affiliation{Key Laboratory of Quark and Lepton Physics (MOE) and Institute of Particle Physics, Central China Normal University, Wuhan 430079, China }

\begin{abstract}
Taking doubly charged particles, positive-negative charge pair production and the effects of volume fluctuations into account, the Poisson baseline of the fluctuations of net-charge is studied. Within the Poisson baseline, the cumulants of net-charge are derived. Comparing to the Skellam baseline of net-charge, we infer that doubly charged particles broaden the distributions of net-charge, while positive-negative charge pairs narrow the distributions. Using the ratios of doubly charged particles and positive-negative charge pairs from neutral resonance decays to the total positive charges from THERMINATOR 2, the first four orders of moments and the corresponding moment products are calculated in the Poisson baseline for Au + Au collisions at $\sqrt{s_{NN}}$ = 200 GeV at RHIC/STAR. We find that the standard deviation is mainly influenced by the resonance decay, while the third and fourth order moments and corresponding moment products are mainly modified and fit the data of RHIC/STAR much better after including the effects of volume fluctuations.

\end{abstract}


\maketitle
\section{Introduction}

The high-order cumulants of conserved charges are suggested as good probes of the quantum chromodynamics (QCD) phase diagram~\cite{Stephanov-prl102,koch}. They are experimentally accessible and theoretically calculable. In theory, non-monotonic behavior and even sign changes can be found in the high-order cumulants~\cite{MCheng-PRD79,Fu weijie-PNJL,PQM model,Stephanov-prl107,Asakawa-prl103}. In experiments, the cumulants of net-proton distributions and net-charge distributions are calculated based on the data taken by the Solenoid Tracker at the Relativistic Heavy Ion Collider (RHIC/STAR) with a wide range of collision energies from $\sqrt{s_{NN}}$ = 7.7 GeV to $\sqrt{s_{NN}}$ = 200 GeV~\cite{Phys. Rev. Lett.112.032302,Phys. Rev. Lett.113.092301}.

Before some interpretations from the results of cumulants measured at RHIC/STAR in terms of QCD critical phenomena, the contributions of non-critical fluctuations from other known physics must be quantified, such as the statistical fluctuations due to finite numbers of produced particles~\cite{Phys. Rep.270.1-141, Phys. Lett. B.724.51-55,Phys. Rev. C.89.014904,Nucl. Phys. A.947.248-259}, global conservation laws in a subsystem~\cite{Phys. Rev. C.87.014901}, volume fluctuations~\cite{initial size 1,initial size 2,initial size 3}, and experimental acceptance cuts~\cite{cut,Phys. Lett. B.738.305-310}. It is also suggested to study the dynamical cumulants, which is the difference of the cumulants calculated from experiments and corresponding statistical fluctuations~\cite{Phys. Rev. C.79.024906,Lizhu-JPG}. Usually, the statistical fluctuations are considered as a baseline.

The cumulants or cumulant ratios of net-protons measured at RHIC/STAR are often compared to a baseline that assumes Poisson and negative binomial statistics. In case of the Poisson statistics, the proton and anti-proton multiplicities are randomly sampled from their mean values, resulting in a Skellam distribution of net-protons~\cite{Phys. Rev. Lett.112.032302}. The baselines based on negative binomial distributions (NBDs) are constructed by using both the measured mean values and variances of the proton and anti-proton~\cite{Phys. Lett. B.724.51-55}.

Turning to the baseline of net-charge, is it still proper to assume a Poisson distribution or NBD for the total positive and negative charges? In fact, in the cases of Poisson and NBD statistics, the particles are all produced independently. This assumption is suitable for the baseline of the proton and antiproton, but not so reasonable for positive and negative charges.

On one hand, a lot of charged particles come from resonance decays~\cite{Eur. Phys. J. C.75.573}. These contribute in two ways. One is from doubly-charged particles, which is included in the data through their decay products. The other is from positive-negative charge pairs generated from the resonance decay. Strong correlations are reserved for the positive-negative charge pairs. On the other hand, quantum effects are more crucial for small masses, such as pions. They cannot be ignored in the studies of net-charge fluctuations. In Ref.~\cite{Nucl. Phys. A.880.48-64}, it was shown that the contribution of quantum effects broadens the distribution of net-charge, just like the doubly charged particles.

In this paper, taking these two aspects of the contribution of resonance decays into account, we study the Poisson statistics with a new assumption. First, the total positive or negative charges are divided into three parts. The first part is from  doubly charged particles, the second part is from  positive-negative charge pair production, and the third part consists of all the rest. The three groups of particles are all assumed to follow Poisson distributions. Under this framework, a Poisson baseline of net-charge fluctuations can be derived.

Under the assumption that the fluctuations of charged particle number and volume are independent, the effects of the volume fluctuations are included in the Poisson baseline. Usually, a Glauber model that includes nuclear geometry and particle production is used to generate the volume or rather participant fluctuations ~\cite{Glauber}. Based on the Glauber model, the distribution of participants is determined by a certain centrality selection. In the case of net-charge distributions measured by the STAR collaboration, however, the contribution of volume fluctuations can be approximately carried out in non-central collisions~\cite{Phys. Rev. C.94.054903}.

The paper is organized as follows. The cumulants of net-charge in the Poisson baseline are derived after taking the doubly charged particles and positive-negative charge pair production into account in Section 2. The effects of volume fluctuations on the cumulants of net-charge are studied in Section 3. In Section 4, comparing with the Skellam baseline of net-charge, the influence on the distributions of net-charge from doubly charged particles is discussed, as is the influence of the positive-negative charge pairs. With the ratios of doubly charged particles and positive-negative charge pairs to the total positive charges from THERMINATOR 2, which simulates Au + Au collisions at $\sqrt{s_{NN}}$ = 200 GeV, the first four orders of moments and corresponding moment products from the Poisson baseline and the Poisson baseline, including the effects of volume fluctuations, are calculated. The results are compared to the Skellam baseline and the data from RHIC/STAR. Finally, we summarize in Section 5.

\section{Framework of the Poisson baseline}

Assuming that the distribution of particle multiplicity follows a Poisson distribution is equivalent to the assumption that the number of particles produced in each event is a discrete random variable $N$, and the probability mass function (PMF) of $N$ is given by,
\begin{flalign}\label{PMF of Po}
 f(k;\lambda)=Pr(N=k)=\frac{\lambda^ke^{-\lambda}}{k!},~ k=0,1,2...,
\end{flalign}
where $\lambda>0$ equals the expected value and also the variance of $N$.

If the number of positive and negative charges is assumed to follow the Poisson distribution directly, as done in Ref.~\cite{Phys. Rev. Lett.113.092301}, then the net-charge will follow the Skellam distribution~\cite{Skellam}. Its cumulants are only determined by the mean value of the positive charge ($M_+$) and negative charge ($M_-$) as follows,
\begin{equation}\label{cumulants of Sk}
 C_{2n-1}^S=M_+ - M_-, ~C_{2n}^S=M_+ + M_-,~ n=1,2,3...，
\end{equation}
where $C_{2n-1}^S$ and $C_{2n}^S$ represent the odd-order and even-order of cumulants, respectively. $C_1^S$ and $C_2^S$ are the mean and variance of net-charge distribution in the Skellam baseline.

In this paper, considering the contribution of resonance decay to the charge distributions, we use six discrete random variables $N_{2+}$ ($N_{2-}$), $N_{p+}$ ($N_{p-}$), and $N_{1+}$ ($N_{1-}$) to represent the numbers of doubly positive (negative) charged particles, singly positive (negative) charged particles from positive-negative charge pair productions, and the rest singly positive (negative) charged particles in each event, respectively. The decay products of the doubly charged particles are assumed to be in the same event, and so are the positive-negative charged particles from pair production. The six discrete random variables are all assumed to follow Poisson distributions.

The positive (negative) charges from the doubly positive (negative) charged particles no longer follow Poisson distributions. Their PMF is as follows,
\begin{equation}\label{PMF of charge2}
 f(2k;2\langle N_{2+} \rangle)=\frac{\langle N_{2+} \rangle^ke^{-\langle N_{2+} \rangle}}{k!},~~~ k=0,1,2...,
\end{equation}
where $\langle N_{2+} \rangle$ represents the mean value of $N_{2+}$. Then $2\langle N_{2+} \rangle$ is the mean value of charges taken by doubly positive charged particles.

As we know, the sum of two Poisson distributions is still a Poisson distribution. Its expected value is the sum of the expected values of the two Poisson distributions. Then the PMF of the total positive charge ($N_+$) is the convolution of PMF of $N_{1+}$, $2N_{2+}$ and $N_{p+}$. It can be written as follows,
\begin{equation}\label{PMF of total charge}
\begin{split}
f(k;\langle N_{+} \rangle)=\sum_{x=-\infty}^{\infty}f(x;2N_{2+})f(k-x;\langle N_{1+}+N_{p+} \rangle)\\
=\sum_{x=-\infty}^{\infty}\frac{\langle N_{2+} \rangle^{x/2}e^{-\langle N_{2+} \rangle}}{(x/2)!}\frac{\langle N_{1+}+N_{p+} \rangle^{k-x}e^{-\langle N_{1+}+N_{p+} \rangle}}{(k-x)!},\\
k=0,1,2...,
\end{split}
\end{equation}

The $n_{th}$-order cumulants ($C_n^{N_{+}}$) of the total charges can be derived from its cumulant generating function (CGF),
\begin{equation}\label{CGF of total charge}
 K_{N_{+}}(t; \langle N_{+}\rangle)=\sum_{k=0}^{\infty}\frac{t^k}{k!}C_k^{N_{+}},
\end{equation}
where $K_{N_{+}}(t; \langle N_{+}\rangle)=\ln G(e^t; \langle N_{+}\rangle)$, and $G(t; \langle N_{+}\rangle)$ is the probability generating function (PGF), i.e.,
\begin{equation}\label{PGF of total charge}
 G(t; \langle N_{+}\rangle)=\sum_{k=0}^{\infty}f(k;\langle N_{+}\rangle)t^{k}.
\end{equation}
So the $n_{th}$-order cumulant of total charge is as follows,
\begin{equation}\label{cumulants of total charge}
 C_n^{N_{+}}=\langle N_{1+}\rangle+\langle N_{p+}\rangle+2^n\langle N_{2+}\rangle, ~~~n=1,2,3...，
\end{equation}
For the cumulants of total negative charges, one can just  replace $\langle N_{1+}\rangle$, $\langle N_{p+}\rangle$ and $\langle N_{2+}\rangle$ with $\langle N_{1-}\rangle$, $\langle N_{p-}\rangle$ and $\langle N_{2-}\rangle$ in Eq.~\eqref{cumulants of total charge}.

Usually, the PMF of difference of two independent random variables is the cross-correlation of their PMFs. One may think that through the cross-correlation of the PMFs of total positive and negative charges, the PMF of the net-charge can be derived. However, this is not the case here. We should be careful when dealing with the net-charge from positive-negative charged pair production.

The positive and negative charges from pair production are all assumed to follow Poisson distributions. Not only are their expected values equal to each other $\langle N_{p+}\rangle=\langle N_{p-}\rangle$, but also they share the same random number sequence in the same order. If $N_{p+}^i$ and $N_{p-}^i$ are used to represent the number of singly positive and negative charged particles from pair production in the $i_{th}$ event, respectively, one can get the following relation,
\begin{equation}\label{pair production}
 N_{p+}^i=N_{p-}^i.
\end{equation}

\begin{table*}
	\caption{\label{comparison} The values of the ratios for the eight centralities in THERMINATOR 2}
	\begin{tabular}{|l|l|l|l|l|l|}
		\hline
		~~~~Centrality&~~$\langle N_{+}\rangle$~~ &~~~~~$r_{2+}$ & ~~~~~$r_{2-}$& ~~~~~$r_{p+}$&~~~~~$r_{p-}$  \\
		\hline
		~~~~~~~0\%-5\%~ &~~~256.859 & ~~1.07311\%~ &~~0.73937\%  ~&~~19.8436\%  ~ &~~19.8562\%  ~\\
		\hline
		~~~~~~5\%-10\%~ &~~~220.962 & ~~1.06932\%  ~ &~~0.734299\%  ~&~~19.8594\%  ~ &~~19.8737\%  ~\\
		\hline
		~~~~~10\%-20\%~ &~~~171.967~ & ~~1.07345\%  ~ &~~0.73921\%  ~ &~~19.8257\%  ~&~~19.8156\%  ~\\
		\hline
		~~~~~20\%-30\%~ &~~~121.545~ & ~~1.07597\%  ~ &~~0.72834\%  ~ &~~19.8376\%  ~&~~19.848\%  ~\\
		\hline
		~~~~~30\%-40\%~ &~~~82.3718~ & ~~1.05898\%  ~ &~~0.72571\%  ~ &~~19.8124\%  ~&~~19.8796\%  ~\\
		\hline
		~~~~~40\%-50\%~ &~~~52.5943~ & ~~1.05536\%  ~ &~~0.733996\%  ~ &~~19.8603\%  ~&~~19.857\%  ~\\
		\hline
		~~~~~50\%-60\%~ &~~~28.3118~ & ~~1.05221\%  ~ &~~0.737008\%  ~ &~~19.861\%  ~&~~19.9299\%  ~\\
		\hline
		~~~~~60\%-70\%~ &~~~9.1492~ & ~~1.05889\%  ~ &~~0.690552\%  ~ &~~20.0105\%  ~&~~20.0573\%  ~\\
		\hline
	\end{tabular}
\end{table*}

This means that the distribution of net-charge ($N_{p+}-N_{p-}$) from pair production is not a Skellam distribution any more, but zero. In fact, it reflects the conservation law of charge, which has an important influence on net-charge fluctuations. If the experimental acceptance can get to the full space and the detection efficiency is one, the event-by-event net-charge fluctuations will be decided only by the initial volume fluctuations or rather participant fluctuations.

Now,  analyzing the total distribution of net-charge of all charged particles, we just need to consider the contribution of singly charged particles, having already considered those from pair production and doubly charged particles. The PMF of the net-charge is the cross-correlation of the PMFs of $N_{1+} + 2N_{2+}$ and $N_{1-} + 2N_{2-}$. Under this case, the corresponding odd-order ($C_{2n-1}^{N}$) and even-order ($C_{2n}^{N}$) cumulants of net-charge have been derived in Ref.~\cite{Phys. Rev. C.89.014904},
\begin{equation}\label{cumulants of total charge 1}
\begin{split}
& C_{2n-1}^{N}=\langle N_{1+}\rangle-\langle N_{1-}\rangle+2^{2n-1}(\langle N_{2+}\rangle-\langle N_{2-}\rangle),\\
& C_{2n}^{N}=\langle N_{1+}\rangle+\langle N_{1-}\rangle+2^{2n}(\langle N_{2+}\rangle+\langle N_{2-}\rangle). \\
&n=1,2,3...
 \end{split}
\end{equation}

\section{Volume fluctuations}

A general expression for the cumulants of net-baryons including the effects of volume fluctuations is derived in Ref.~\cite{Phys. Rev. C.88.034911}, under the assumption that the fluctuations of the baryon number and volume are independent. The expression is suitable for an arbitrary probability distribution for the fluctuations of net-baryon number as well as for the fluctuations of the volume. If the fluctuations of charged particle number and volume are assumed to be independent, the same expression can be obtained for the cumulants of net-charge.

Let us use $c_{n}=C_{n}^{N}/V$ to represent the $n_{th}$-order reduced cumulants of net-charge, corresponding to the net-charge number fluctuations per unit volume in a fixed volume $V$. If fluctuations of volume are allowed, the first four orders of reduced cumulants $v_{n}$ of volume fluctuations are as follows,
\begin{equation}\label{reduced cumulants of volume}
\begin{split}
&v_{1}=\frac{\langle V\rangle}{\langle V\rangle}=1, \quad v_{2}=\frac{\langle(\delta V)^2\rangle}{\langle V\rangle},\\
&v_{3}=\frac{\langle(\delta V)^3\rangle}{\langle V\rangle},\quad v_{4}=\frac{\langle(\delta V)^4\rangle-3\langle(\delta V)^2\rangle^2}{\langle V\rangle},
\end{split}
\end{equation}
where $\delta V=V-\langle V\rangle$.
Then according to Eq.~(7) in Ref.~\cite{Phys. Rev. C.88.034911}, the first four orders of reduced cumulants of net-charge including the effects of volume fluctuations are as follows,
\begin{equation}\label{reduced cumulants of net-charge}
\begin{split}
&\kappa_1=c_1, \quad \kappa_2=c_2+{c_1}^2v_2,\\
&\kappa_3=c_3+3{c_2}{c_1}v_2+{c_1}^3v_3,\\
&\kappa_4=c_4+(4{c_3}{c_1}+3{c_2}^2)v_2+6{c_2}{c_1}^2v_3+{c_1}^4v_4.
\end{split}
\end{equation}
For a detailed derivation, please see Section II in Ref.~\cite{Phys. Rev. C.88.034911}, replacing the net-baryon number with the net-charge number.

In this paper, the volume is regarded as the number of participants directly. Based on the Glauber model, the distribution of participants is determined by a certain centrality selection. In non-central heavy-ion collisions, however, the second to fourth orders of reduced cumulants of volume fluctuations can be approximated with $v_2=1$, $v_3=2$, $v_4=6$~\cite{Phys. Rev. C.94.054903}. Then the reduced cumulants of net-charge including the effects of volume fluctuations in Eq.~\eqref{reduced cumulants of net-charge} are simplified to
\begin{equation}\label{simplified cumulants of net-charge}
\begin{split}
&\kappa_1=c_1=\frac{C_1^N}{\langle N_{part}\rangle}, \\
&\kappa_2=c_2+{c_1}^2=\frac{C_2^N}{\langle N_{part}\rangle}+\frac{{C_1^N}^2}{{\langle N_{part}\rangle}^2},\\
&\kappa_3=c_3+3{c_2}{c_1}+2{c_1}^3=\frac{C_3^N}{\langle N_{part}\rangle}+\frac{3C_2^NC_1^N}{\langle N_{part}\rangle^2}+\frac{2{C_1^N}^3}{\langle N_{part}\rangle^3},\\
&\kappa_4=c_4+(4{c_3}{c_1}+3{c_2}^2)+12{c_2}{c_1}^2+6{c_1}^4\\
&=\frac{C_4^N}{\langle N_{part}\rangle}+\frac{4C_3^NC_1^N+3{C_2^N}^2}{\langle N_{part}\rangle^2}+\frac{12C_2^N{C_1^N}^2}{\langle N_{part}\rangle^3}+\frac{6{C_1^N}^4}{\langle N_{part}\rangle^4}.
\end{split}
\end{equation}
The corresponding cumulants $C_n^V$ of net-charge including the effects of volume fluctuations are the reduced cumulants multiplying $\langle N_{part}\rangle$,
\begin{equation}\label{simplified cumulants of net-charge}
\begin{split}
&C_1^V=\langle N_{part}\rangle\kappa_1, \quad C_2^V=\langle N_{part}\rangle\kappa_2,\\
&C_3^V=\langle N_{part}\rangle\kappa_3, \quad C_4^V=\langle N_{part}\rangle\kappa_4.
\end{split}
\end{equation}

\begin{figure*}
	\includegraphics[width=0.246\textwidth]{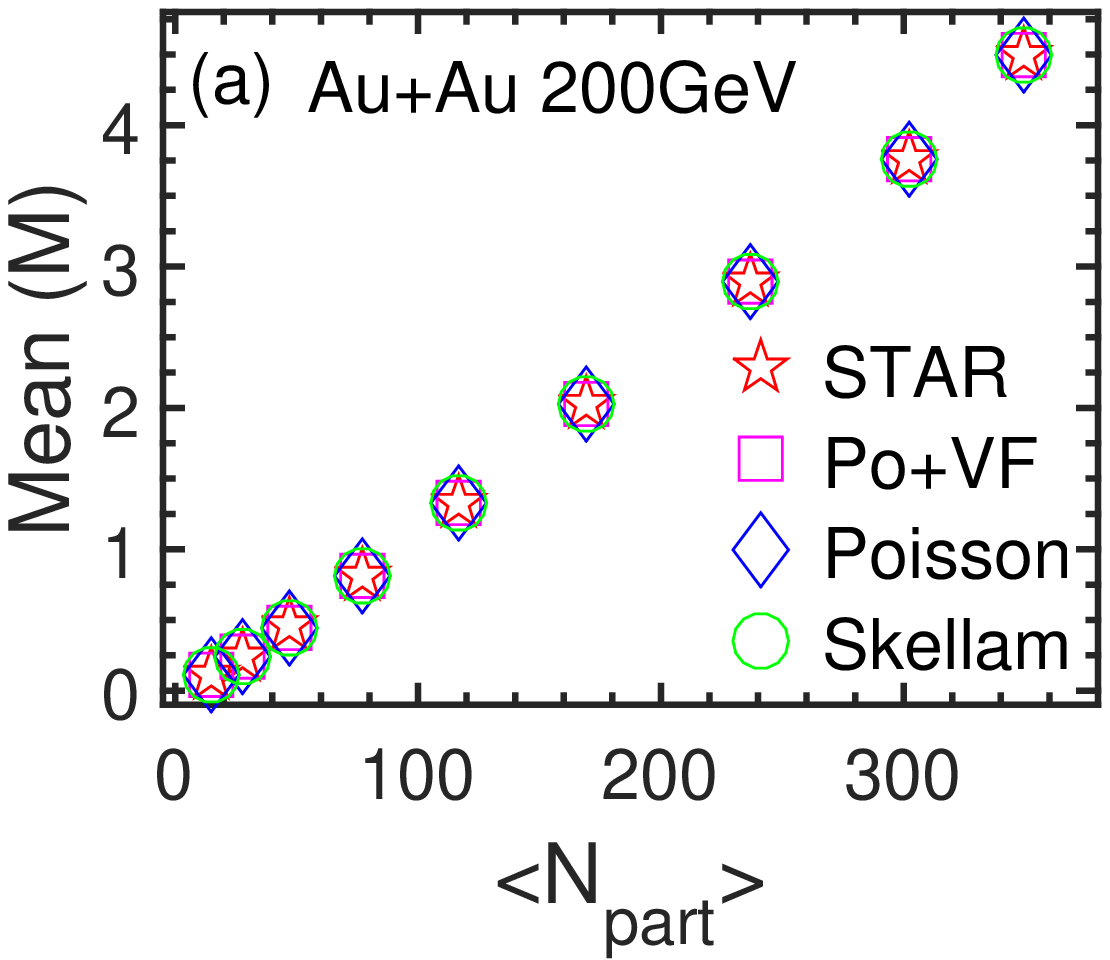}
	\includegraphics[width=0.246\textwidth]{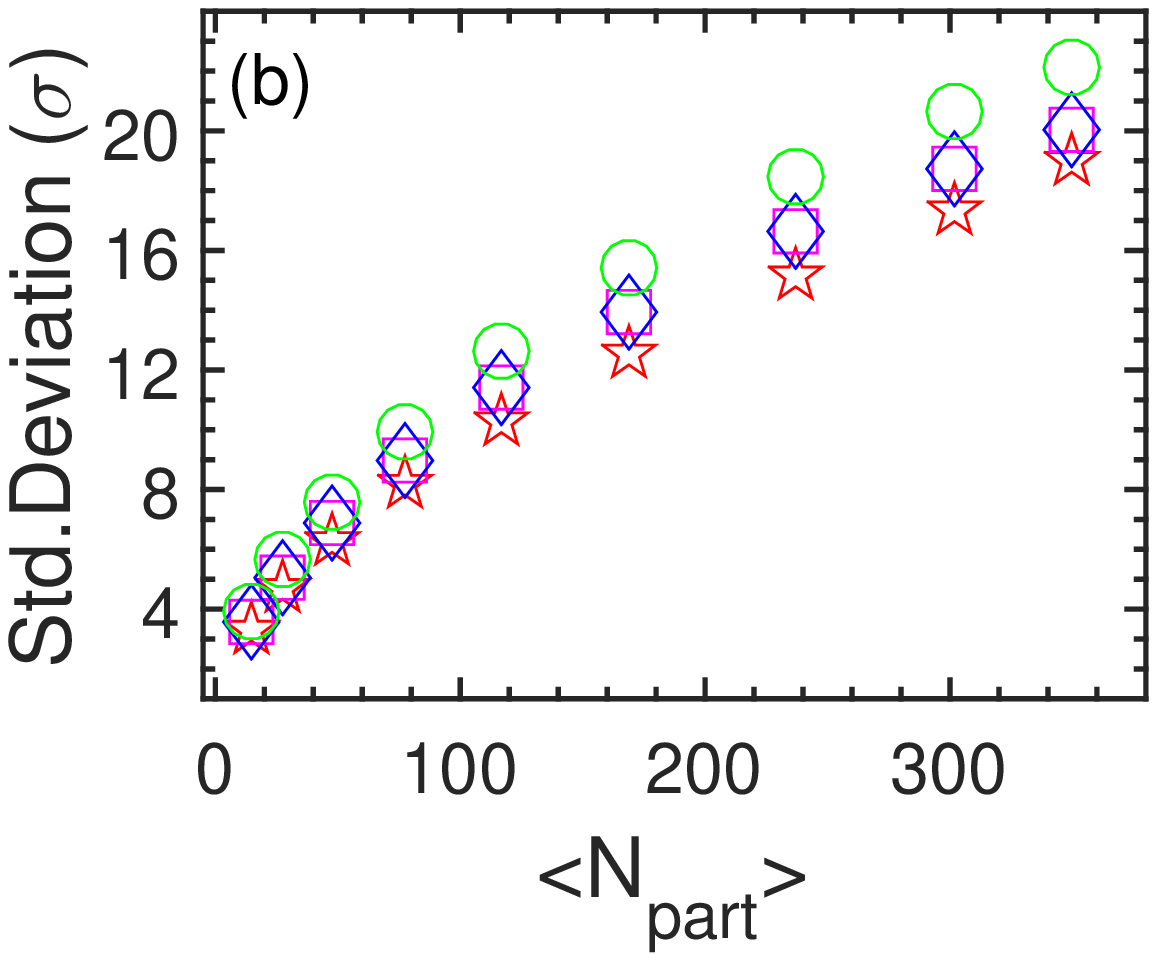}
	\includegraphics[width=0.246\textwidth]{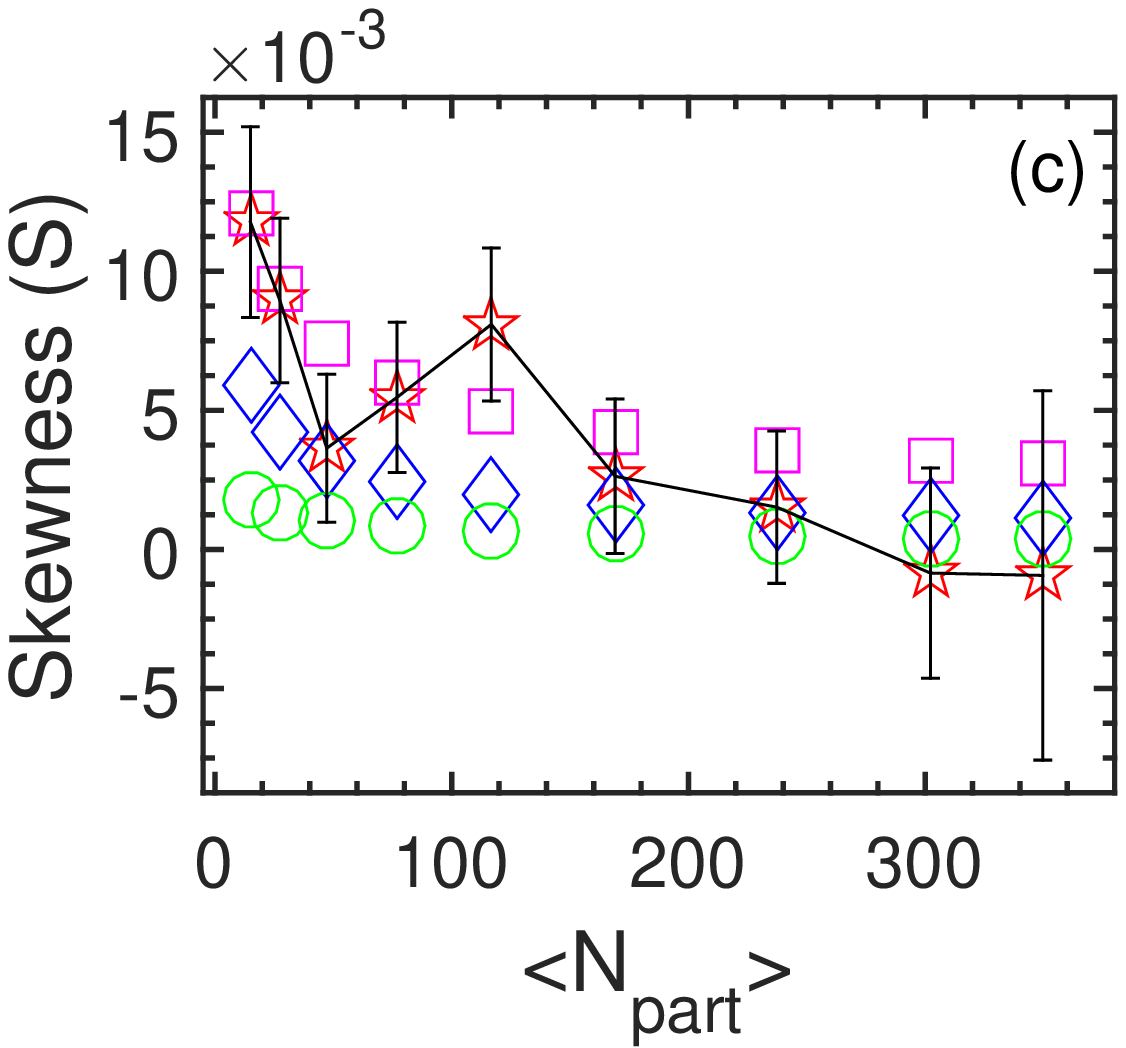}
	\includegraphics[width=0.246\textwidth]{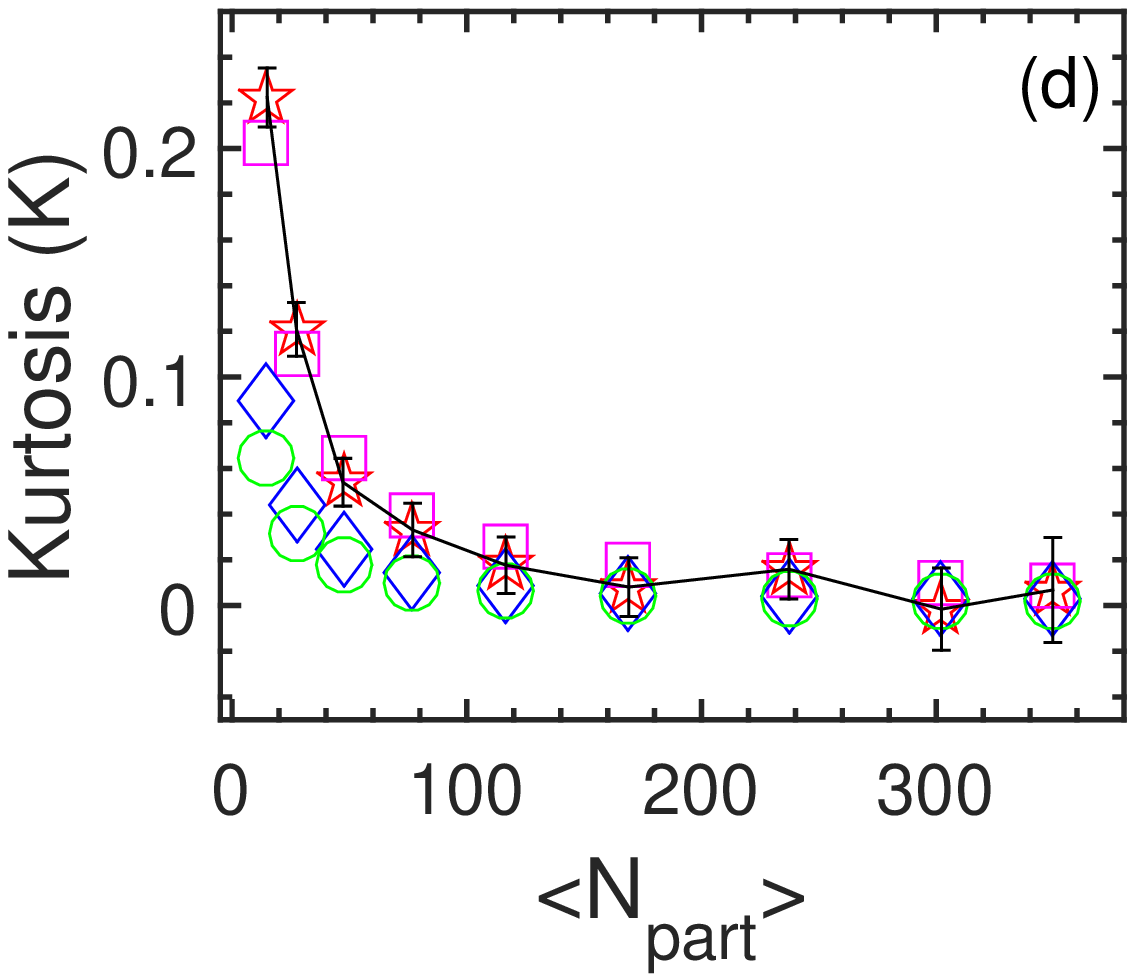}
	\caption{\label{Fig. 1}(Color online). The centrality dependence of (a) mean (M), (b) standard deviation ($\sigma$), (c) skewness (S), and (d) kurtosis (K) of the net-charge distributions from RHIC/STAR (red stars), Poisson baseline including effects of volume fluctuations (pink squares), Poisson baseline (blue diamonds) and Skellam baseline (green circles) with ratios $r_p=19.86\%$, $r_{2+}=1.06\%$ and $r_{2-}=0.73\%$ for Au + Au collisions at $\sqrt{s_{NN}}$ = 200 GeV.}
\end{figure*}

\section{Moments of net-charge from the Poisson baseline including volume fluctuations}

For the Poisson baseline, using the relations $\langle N_{+}\rangle=\langle N_{1+}\rangle + 2\langle N_{2+}\rangle+\langle N_{p+}\rangle$ and $\langle N_{-}\rangle=\langle N_{1-}\rangle + 2\langle N_{2-}\rangle+\langle N_{p-}\rangle$, the first four orders of cumulants of net-charge are as follows,
\begin{equation}\label{cumulants of total charge 2}
\begin{split}
 &C_1^N=\langle N_+\rangle-\langle N_-\rangle,\\
 &C_2^N=\langle N_{+}\rangle + \langle N_{-}\rangle +2(\langle N_{2+}\rangle+\langle N_{2-}\rangle)-(\langle N_{p+}\rangle + \langle N_{p-}\rangle), \\
&C_3^N=\langle N_{+}\rangle-\langle N_{-}\rangle+6(\langle N_{2+}\rangle-\langle N_{2-}\rangle),\\
 &C_4^N=\langle N_{+}\rangle+\langle N_{-}\rangle+14(\langle N_{2+}\rangle+\langle N_{2-}\rangle)-(\langle N_{p+}\rangle + \langle N_{p-}\rangle),
 \end{split}
\end{equation}
where $\langle N_{+}\rangle$ and $\langle N_{-}\rangle$ are the mean values of the total positive and negative charges, respectively.

Comparing $C_2^N$ in Eq.~\eqref{cumulants of total charge 2} and $C_2^S$ in Eq.~\eqref{cumulants of Sk}, it is clear that the doubly charged particles broaden the distribution of net-charge in the Skellam baseline, while the particles from pair production narrow the distribution.

Supposing that the ratio of singly charged particles from pair production to the total positive charges is $r_p=r_{p+}=\langle N_{p+}\rangle/\langle N_+\rangle=r_{p-}=\langle N_{p-}\rangle/\langle N_+\rangle$. Similarly, the ratios of doubly positive and negative charges are $r_{2+}=\langle N_{2+}\rangle/\langle N_+\rangle $, and $r_{2-}=\langle N_{2-}\rangle/\langle N_+\rangle $, respectively. The cumulants in Eq.~\eqref{cumulants of total charge 2} can be written as follows,
\begin{equation}\label{cumulants of total charge 3}
\begin{split}
& C_1^N=\langle N_+\rangle-\langle N_-\rangle,\\
&C_2^N=\langle N_{+}\rangle + \langle N_{-}\rangle+2(r_{2+}+r_{2-}-r_p)\langle N_{+}\rangle,\\
 &C_3^N=\langle N_{+}\rangle-\langle N_{-}\rangle+6(r_{2+}-r_{2-})\langle N_{+}\rangle,\\
&C_4^N=\langle N_{+}\rangle+\langle N_{-}\rangle+(14r_{2+}+14r_{2-}-2r_p)\langle N_{+}\rangle.
\end{split}
\end{equation}

From the expression of $C_2^N$ in Eq.~\eqref{cumulants of total charge 3}, it is clear that it is decided by the ratios $r_{2+}$, $r_{2-}$ and $r_p$ that the distribution of net-charge is wider or narrower in the Poisson baseline than the Skellam baseline. Doubly charged particles are mainly $\Delta^{++}$ and its anti-particle $\Delta^{--}$. Their ratios are not big. However, there are lots of charges from pair production in neutral and also higher-mass resonance decays, whose effects have been studied in a hadron resonance gas model~\cite{Phys. Rev. C.94.014905}.

Through several simple assumptions, the formulas of the cumulants including the resonance decay effects have been derived. It can be inferred through Eq.~\eqref{cumulants of total charge 3} that there is great change in the cumulant after considering the effects of resonance decay.

Assuming that the hadronic matter reaches thermal equilibrium and undergoes rapid expansion leads to a remarkably good description of the ratios of particle abundances measured in heavy-ion experiments~\cite{particle abundance 1,particle abundance 2,particle abundance 3}. In the THERMINATOR 2 model, simulating Au + Au collisions at $\sqrt{s_{NN}}$ = 200 GeV~\cite{Comput. Phys. Commun.183.746-773}, the resonance decay is implemented. Considering the doubly charged $\Delta^{++}$ and $\Delta^{--}$, we calculated the ratios $r_{2+}$ and $r_{2-}$ in THERMINATOR 2 for the eight centralities available in this model. Taking the seven kinds of neutral particles, $\rho^0$, $\eta^0$, $K^0$, $\overline{K}^0$, $\omega$, $K^{*0}$ and $\overline{K}^{*0}$ into account, $r_{p+}$ and $r_{p-}$ for the eight centralities are obtained. The positive and negative charged particle multiplicities within pseudorapidity $\eta$ window of $|\eta|<0.5$ and the transverse momentum $p_T$ range $0.2<p_T<2$ GeV/$c$ are taken into account, in the same way as in Ref.~\cite{Phys. Rev. Lett.113.092301}.

\begin{figure*}
	\includegraphics[width=0.32\textwidth]{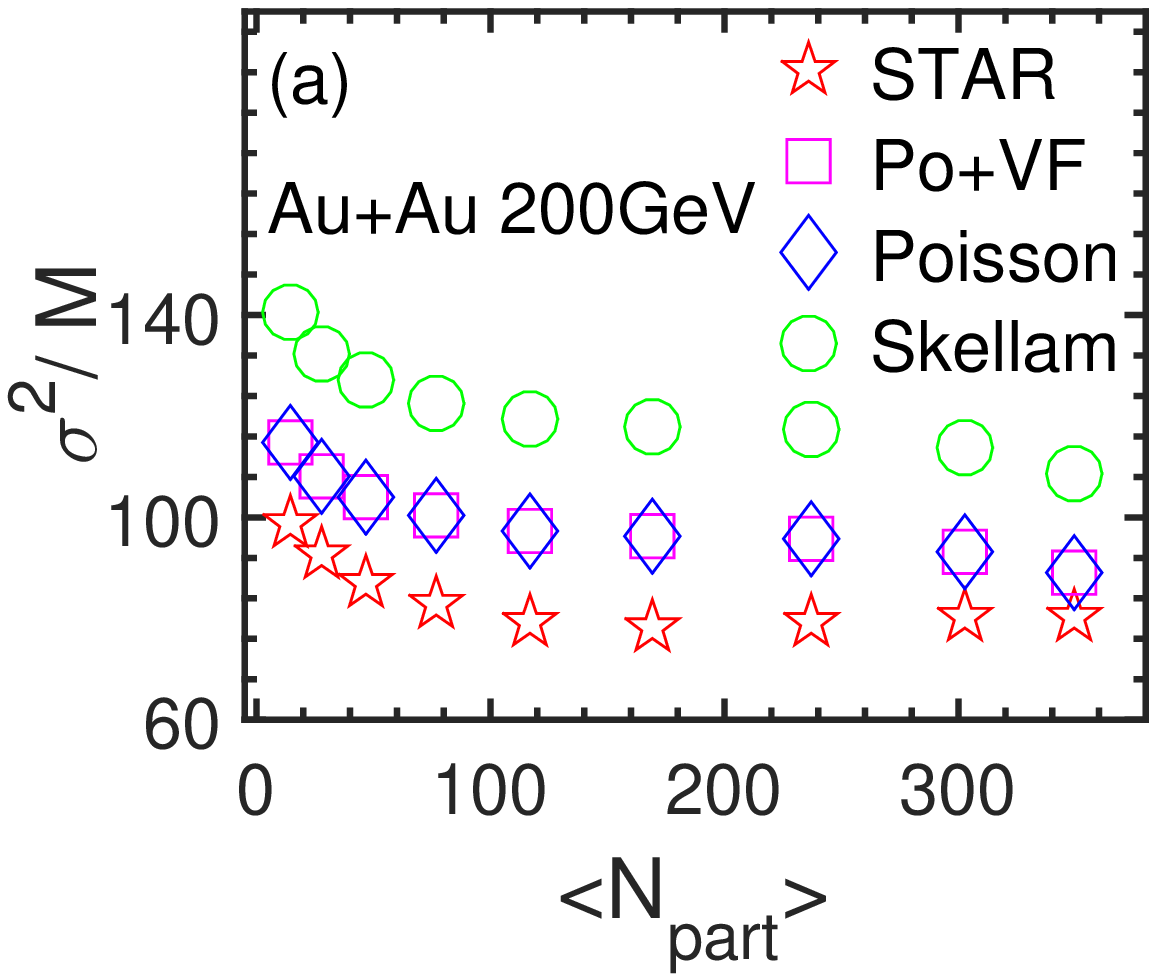}
	\includegraphics[width=0.32\textwidth]{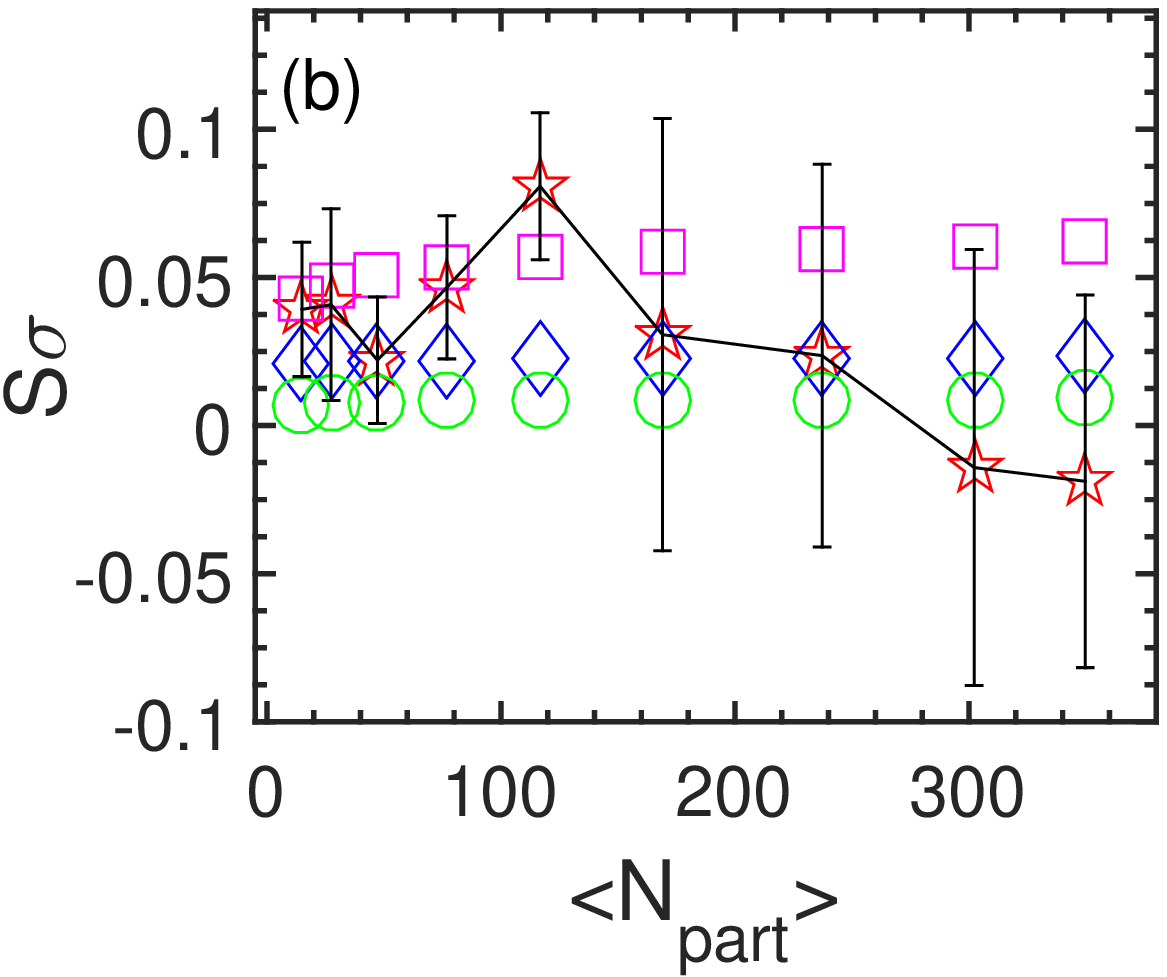}
	\includegraphics[width=0.32\textwidth]{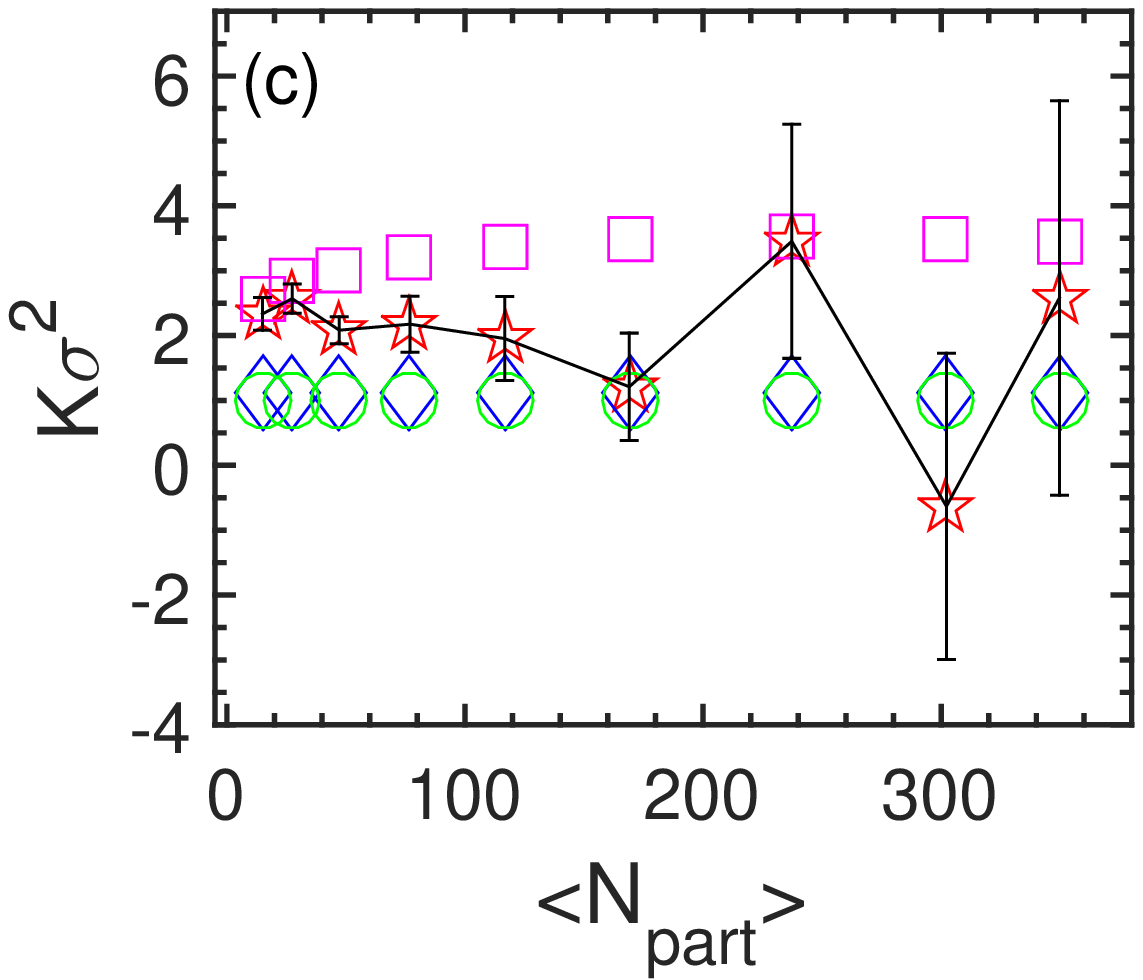}
	\caption{\label{Fig. 2}(Color online). The centrality dependence of the moment products $\sigma^2/M$, $S\sigma$, and $K\sigma^2$ of net-charge distributions from RHIC/STAR (red stars), Poisson baseline including effects of volume fluctuations (pink squares), Poisson baseline (blue diamonds) and Skellam baseline (green circles) with ratios $r_p=19.86\%$, $r_{2+}=1.06\%$ and $r_{2-}=0.73\%$ for Au + Au collisions at $\sqrt{s_{NN}}$ = 200 GeV.}
\end{figure*}

The mean values of the multiplicities of total positive charges $\langle N_{+}\rangle$, and the values of $r_{2+}$, $r_{2-}$, $r_{p+}$ and $r_{p-}$ for the eight centralities are shown in Table 1. From the results, we find the values of $r_{2+}$ are around $1.06\%$. The difference is less than $0.02\%$ for each centrality. The values of $r_{2-}$ are around $0.73\%$. The difference is less than $0.01\%$ except for the centrality 60\%-70\%. The same happens for $r_{p+}$ and $r_{p-}$. In fact, the values of each ratio are almost the same for all of the centralities, with no change as the centralities vary. The slight difference, less than 0.1\%, between $r_{p+}$ and $r_{p-}$ for each centrality could be caused by the cuts of transverse momentum and pseudorapidity. Because there is so little difference, it is acceptable to choose the same value of the ratio for all of the centralities. It is also demonstrated that it is fine to set $r_{p+}=r_{p-}$.

Taking $r_{2+}=1.06\%$, $r_{2-}=0.73\%$ and $r_{p}=r_{p+}=r_{p-}=19.86\%$ in the nine centralities of Au + Au collisions at $\sqrt{s_{NN}}$ = 200 GeV at RHIC/STAR, the centrality dependence of the mean ($M=C_1$, $C_n$ representing the $n_{th}$-order cumulant) in the Poisson baseline and including the effects of volume fluctuations are shown by the blue diamonds and pink squares in Fig. 1(a), respectively. The corresponding standard deviation ($\sigma=\sqrt{C_2}$), skewness ($S=C_3/{C_2}^{1.5}$) and kurtosis ($K=C_4/{C_2}^2$) are shown in Fig. 1(b), 1(c) and 1(d), respectively. The red stars are data from RHIC/STAR~\cite{Phys. Rev. Lett.113.092301}, and the green circles are the baseline from the Skellam distribution. The centrality is represented by the average number of participating nucleons $\langle N_{part}\rangle$. In Fig. 1(a), there is no doubt that the mean value from each baseline is the same as the data.

In Fig. 1(b), $\sigma$ from the Skellam baseline is systematically bigger than data at all centralities. After considering the doubly charged particles and particles from pair production, the Poisson baseline gets closer to the data. After including the effects of volume fluctuations, the Poisson baseline of $\sigma$ is almost unchanged.

For skewness and kurtosis in Fig. 1(c) and Fig. 1(d), the results from the Skellam baseline are systematically smaller than the data, while the Poisson baseline is closer to the data. After including the effects of volume fluctuations, the Poisson baselines of $S$ and $K$ change a lot and are very close to the data.

For standard deviation, the differences between the Skellam and Poisson baselines increase going from peripheral collisions to central collisions, as do the differences between data and the baselines. For skewness and kurtosis, the opposite is true. The differences become smaller going from peripheral collisions to central collisions.

To explore  the Poisson baseline and effects of volume fluctuations of net-charge distributions in depth, we further compare the moment products $\sigma^2/M$, $S\sigma$, and $K\sigma^2$ from the three different kinds of baselines and data, as shown in Fig.~2. Differences between the Poisson baseline and the Skellam baseline decrease with the increase of the order of the moments. After including the effects of volume fluctuations, $\sigma^2/M$ is almost unchanged. $S\sigma$ and $K\sigma^2$ change a lot.

In short, the standard deviation is more sensitive to the resonance decay and less sensitive to the volume fluctuations. The opposite is true for the third and fourth order moments or moment products. These results are similar to the results from a Monte Carlo hadron resonance gas model in Ref.~\cite{initial size 3}.

There are still large differences between the baselines and data for $\sigma^2/M$. As well as the strong correlation caused by the doubly charged particles and positive-negative charge pairs from neutral resonance decays, there should still exist some other strong correlations between the charged particles, which causes the distribution of net-charge to be narrower in the data than in the Poisson baseline. If higher-mass resonance decays are included, the value of $r_{p}$ should be larger and the Poisson baseline may be closer to the data.

\section{Summary}

Taking doubly charged particles and positive-negative charge pair production from resonance decays into account, and assuming the numbers of doubly positive (negative) charged particles, singly positive (negative) charged particles from positive-negative charge pair productions, and the remaining singly positive (negative) charged particles all follow Poisson distributions, the cumulants of net-charge distributions in the Poisson baseline are derived. The effects of volume fluctuations are also studied under the assumption that the fluctuations of charged particle number and volume are independent.

Comparing with the Skellam distribution, we found that doubly charged particles broaden the distribution of net-charge, while positive-negative charge pairs narrow the distribution.

Through a THERMINATOR 2 simulation for Au + Au collisions at $\sqrt{s_{NN}}$ = 200 GeV, the ratios of doubly positive (negative) charged particles or positive-negative charge pairs from neutral resonance decay to the total positive charges were simulated. Using these ratios, the mean, standard deviation, skewness and kurtosis, and their products from the Poisson baseline, including the effects of volume fluctuations, were calculated for Au + Au collisions at $\sqrt{s_{NN}}$ = 200 GeV from RHIC/STAR. We found that the Poisson baseline, especially after including the effects of volume fluctuations, is closer to the data than the Skellam baseline. The standard deviation is more sensitive to the resonance decay and less sensitive to the volume fluctuations, while the opposite is true for the third and fourth order moments or moment products.

There are still large differences in $\sigma^2/M$ between the Poisson baseline and data. There should still exist some other correlations between the charged particles, which cause the distribution of net-charge to be narrower in the data than in the Poisson baseline.

This work is supported by the NSFC of China under Grants No. 11647093, 11405088 and 11521064, Fund Project of Chengdu Technological University under Grant No. 2016RC004, the Major State Basic Research Development Program of China under Grant No. 2014CB845402.

\ed